\documentclass[12pt]{article}
\usepackage[margin=2.5cm]{geometry}
\usepackage{amsmath, amssymb}
\usepackage{color}
\usepackage{float}
\usepackage{graphicx}
\usepackage{subcaption,adjustbox}
\renewcommand{\Re}{\mathrm{Re}}
\renewcommand{\Im}{\mathrm{Im}}

\begin{document}

\title{Softly-broken $A_4$ or $S_4$ 3HDMs with stable states}

\author{
Ivo de Medeiros Varzielas \footnote{ivo.de@udo.edu},
Diogo Ivo \footnote{diogo.ivo@tecnico.ulisboa.pt}\\
CFTP, Departamento de Física,\\
Instituto Superior T\'{e}cnico, Universidade de Lisboa, \\ Avenida Rovisco Pais 1, 1049 Lisboa, Portugal \\[5pt]
}

\maketitle

\begin{abstract}
We study multi-Higgs models with a triplet of $A_4$ or $S_4$ symmetry that is softly-broken. Focusing on parameters that softly break the respective symmetries but preserve the directions of the vacuum expectation values of the triplet we analyse the masses of physical states and discuss their decays. In contrast with previously studied symmetries, we find some cases with stable states that are protected by residual symmetries left unbroken by both the soft-breaking parameters and by the vacuum.
\end{abstract}

\section{Introduction}

A very natural extension of the Standard Model (SM) is adding scalars to the field content (see e.g. the reviews \cite{Branco:2011iw,Kanemura:2014bqa,Ivanov:2017dad,Arcadi:2019lka}). The class of models where extra $SU(2)$ doublets are introduced, Multi-Higgs Doublet Models with N doublets (NHDM) is rather appealing as it enables possibilities such as scalar CP violation and dark matter candidates.
The 2HDM has been widely studied and 3HDMs, introduced by \cite{Weinberg:1976hu}, are being actively explored. The most general 3HDM has the issue of proliferation of physical parameters, so the typical approach is to consider additional symmetry that makes the model more predictive. The discrete symmetries that can be considered in the 3HDM have been classified in \cite{Ivanov:2012fp} (see also \cite{Darvishi:2019dbh,Darvishi:2021txa}), and their CP properties studied in \cite{Nishi:2006tg, deMedeirosVarzielas:2017ote, Ivanov:2018ime, Ivanov:2019kyh, deMedeirosVarzielas:2019rrp}, with the minima studied in \cite{Degee:2012sk, deMedeirosVarzielas:2017glw}.

With large non-Abelian symmetries such as $\Sigma(36)$, the predictive power is such that the model is not viable, so an interesting possibility is that the large symmetry is present but softly-broken.
This idea was discussed in
\cite{deMedeirosVarzielas:2021zqs}, and methods were developed to apply it to general 3HDMs, and exemplified through the study of the $\Sigma(36)$ case. In the present work, we apply the same methodology to softly-broken $A_4$ and $S_4$ symmetric 3HDMs. In the potentials considered here, we find some qualitatively different results, of particular relevance that there are situations where there are unbroken residual symmetries that stabilize mass eigenstates against decay - which enables these as Dark Matter (DM) candidates.

In Section \ref{sec:A4S4} we review the $A_4$ and $S_4$ potentials, considering the mass spectrum for the possible vacuum expectation values (vevs). In Section \ref{sec:softly} we discuss the parameters that softly break the symmetry, the deviations induced in the masses, and analyse residual symmetries together with associated stablet states. In Section \ref{sec:con} we conclude.

\section{$A_4$ and $S_4$ symmetric 3HDMs \label{sec:A4S4}}

\subsection{Potentials}
In this work we focus on the $S_4$ and $A_4$ symmetric potentials, due to the close relationship between these groups.

This fact leads to both potentials being written in a common form, according to the notation in \cite{Degee:2012sk}

\begin{align}
  V = & - \frac{M_0}{\sqrt{3}} \left(\phi_1 ^ \dagger \phi_1 + \phi_2 ^ \dagger \phi_2 + \phi_3 ^ \dagger \phi_3\right) + \frac{\Lambda_0}{3}  \left(\phi_1 ^ \dagger \phi_1 + \phi_2 ^ \dagger \phi_2 + \phi_3 ^ \dagger \phi_3\right) ^ 2 \notag \\
  & + \Lambda_1 \left( (\Re\phi_1^\dagger\phi_2)^2 + (\Re\phi_2^\dagger\phi_3)^2 + (\Re\phi_3^\dagger\phi_1)^2 \right)\notag + \Lambda_2 \left( (\Im\phi_1^\dagger\phi_2)^2 + (\Im\phi_2^\dagger\phi_3)^2 + (\Im\phi_3^\dagger\phi_1)^2 \right)\\
  & + \frac{\Lambda_3}{3} \left( (\phi_1 ^ \dagger \phi_1) ^ 2 + (\phi_2 ^ \dagger \phi_2) ^ 2 + (\phi_3 ^ \dagger \phi_3) ^ 2  - (\phi_1^\dagger \phi_1) (\phi_2^\dagger \phi_2) - (\phi_2^\dagger \phi_2) (\phi_3^\dagger \phi_3) - (\phi_3^\dagger \phi_3) (\phi_1^\dagger \phi_1)\right) \notag \\
  & + \Lambda_4 \left( (\Re\phi_1^\dagger\phi_2)(\Im\phi_1^\dagger\phi_2) +  (\Re\phi_2^\dagger\phi_3)(\Im\phi_2^\dagger\phi_3) + (\Re\phi_3^\dagger\phi_1)(\Im\phi_3^\dagger\phi_1) \right) \,.
  \label{eqn:potential}
\end{align}

To obtain the $S_4$ invariant potential one just needs to set $\Lambda_4 = 0$, meaning that any results obtained for the $A_4$ invariant model carry over to $S_4$ by means of this substitution, further justifying our dual analysis. As such, the $A_4$ and $S_4$ models have 6 and 5 free parameters, respectively.

These models have 4 different types of vev directions that can lead to a global minimum, \cite{Ivanov:2012fp, deMedeirosVarzielas:2017glw}

\begin{align}
\label{eq::symmvevs}
  (1,0,0) \qquad \qquad (1,1,1) \qquad \qquad (1,e^{i\alpha},0) \qquad \qquad (1,\omega,\omega^2) \,,
\end{align}

\noindent with $\alpha$ fixed by the relations

\begin{equation}
  \sin 2\alpha= - \frac{\Lambda_4}{\sqrt{\left( \Lambda_1 - \Lambda_2 \right)^2 + \Lambda_4^2}} \quad\quad  \quad \cos 2\alpha = - \frac{\Lambda_1 - \Lambda_2}{\sqrt{\left( \Lambda_1 - \Lambda_2 \right)^2 + \Lambda_4^2}} \,,
  \label{eqn:alpha}
\end{equation}

\noindent The presence of these 4 classes manifests itself in 4 possibilities for the predictions of these models, with a set of inequalities on the $\Lambda_i$ selecting which one is the global minimum. These inequalities are also important to take into consideration when taking the limit of $\Lambda_4=0$ in \eqref{eqn:alpha}, since they enforce that $\Lambda_1 - \Lambda_2 > 0$ and thus select $\alpha=\pi/2$ as the global minimum for $S_4$.

\subsection{Fully-symmetric mass spectra}

The spectra for $A_4$ and $S_4$ have been studied in \cite{Degee:2012sk}, and here we summarize those results. Throughout this work, $H_i$ and $h_i$ refer to the heaviest and lightest pair of additional scalars, while $H_{i}^\pm$ refers to the lightest and heaviest of the charged pairs (the index $i$ can take the values $i=1,2$). The Higgs-like boson, $H_{SM}$, always has a mass given by $m_{H_{SM}}^2 = \frac{2}{\sqrt{3}}M_0$, and is therefore omitted. In all cases, $v^2 = v_1^2 + v_2^2 + v_3^2$.

\subsubsection*{$\mathbf{(1, 0, 0)}$:}

For this alignment $\mathbf{(1, 0, 0)}$ to be the global minimum, one must have that
\begin{align}
    \Lambda_3 < 0 & &  \Lambda_0 > |\Lambda_3| > -\Lambda_2, -\Lambda_1 & & \Lambda_4^2 < 4(\Lambda_1 + |\Lambda_3|)(\Lambda_2 + |\Lambda_3|).
\end{align}
This guarantees this is the favored direction, and the magnitude of the vev is then given by
\begin{equation}
    v^2 = \frac{\sqrt{3}M_0}{\Lambda_0 - |\Lambda_3|}.
\end{equation}
The spectra is then expressed as
\begin{align}
m_{h_i}^2 & = \frac{v^2}{4} \left(\Lambda_1 + \Lambda_2 + 2 |\Lambda_3| - \sqrt{(\Lambda_1 - \Lambda_2)^2 + \Lambda_4^2} \right) \notag \qquad \textrm{(double degenerate)}\\
m_{H_i}^2 & = \frac{v^2}{4} \left(\Lambda_1 + \Lambda_2 + 2 |\Lambda_3| + \sqrt{(\Lambda_1 - \Lambda_2)^2 + \Lambda_4^2} \right) \notag  \qquad \textrm{(double degenerate)}\\
m_{H^\pm_i}^2 & = \frac{v^2}{2} |\Lambda_3| \qquad \textrm{(double degenerate)}.
\end{align}

\subsubsection*{$\mathbf{(1, 1, 1)}$:}

In the region of parameter space given by the inequalities
\begin{align}
    \Lambda_1 < 0 & &  \Lambda_0 > |\Lambda_1| > -\Lambda_2, -\Lambda_3 & & \Lambda_4^2 < 12\Lambda_1^2 & & \Lambda_4^2 < 2(\Lambda_3 + |\Lambda_1|)(\Lambda_2 + |\Lambda_1|),
\end{align}
we have the alignment $\mathbf{(1, 1, 1)}$ as the global minimum, with magnitude
\begin{equation}
    v^2 = \frac{\sqrt{3}M_0}{\Lambda_0 - |\Lambda_1|}.
\end{equation}
The spectra has less degeneracies (note the charged Higgses)
\begin{align}
    m_{h_i}^2 & = \frac{v^2}{12} \left( 5 |\Lambda_1| + 3 \Lambda_2 + 2 \Lambda_3 - \sqrt{(|\Lambda_1| + 3 \Lambda_2 - 2 \Lambda_3)^2 + 12 \Lambda_4^2} \right) \notag \qquad \textrm{(double degenerate)} \\
    m_{H_i}^2 & = \frac{v^2}{12} \left( 5 |\Lambda_1| + 3 \Lambda_2 + 2 \Lambda_3 + \sqrt{(|\Lambda_1| + 3 \Lambda_2 - 2 \Lambda_3)^2 + 12 \Lambda_4^2} \right) \notag \qquad \textrm{(double degenerate)} \\
    m_{H^\pm_i}^2 & = v^2 \left(\frac{|\Lambda_1|}{2} \pm \frac{\Lambda_4}{4 \sqrt{3}}\right) .
\end{align}

\subsubsection*{$\mathbf{(1, e^{i\alpha}, 0)}$:}

When the parameters of the potential obey the inequalities
\begin{align}
    \Lambda_2 < 0 & &  |\Lambda_2| > |\Lambda_3| & & \Lambda_1> \Lambda_3 & & 4\Lambda_0 + 3 \Lambda_3 > 3|\Lambda_2|,
\end{align}
we obtain $(1, e^{i\alpha}, 0)$ as the alignment with magnitude
\begin{equation}
    v^2 = \frac{4 \sqrt{3} M_0}{4 \Lambda_0 + \Lambda_3 - 3 \Tilde{\Lambda}},
\end{equation}
where we have defined
\begin{equation}
    \Tilde{\Lambda} = \frac{1}{2} \left( \sqrt{(\Lambda_1 - \Lambda_2)^2 + \Lambda_4^2} - (\Lambda_1 - \Lambda_2)\right).
\end{equation}

The corresponding spectra is
\begin{align}
    m_{H_i}^2 & = \frac{v^2}{4} \left( - (\Lambda_3 + \Tilde{\Lambda}) + ( 1 \pm \cos 3\alpha ) \sqrt{(\Lambda_1 - \Lambda_2 )^2 + \Lambda_4^2} \right) \,, \notag \\
    & = \frac{v^2}{2}(\Lambda_1 + \Lambda_2 + 2 \Tilde{\Lambda}), \qquad \frac{v^2}{2} ( \Lambda_3 + \Tilde{\Lambda}) \notag \\
    m_{H^\pm_i}^2 & = v^2 \frac{\Tilde{\Lambda}}{2}, \qquad v^2 \frac{\Tilde{\Lambda} - \Lambda_3}{4} \,,
\end{align}
where in this case there are no degeneracies and the four $m_{H_i}^2$ are separated by commas.

\subsubsection*{$\mathbf{(1, \omega, \omega^2)}$:}

In the region of parameter space where
\begin{align}
    \Lambda_2 < 0 & &  |\Lambda_2| > |\Lambda_1| & & \Lambda_3> \Lambda_1 & & 4\Lambda_0 + 3 \Lambda_1 > 3|\Lambda_2|,
\end{align}
$\mathbf{(1, \omega, \omega^2)}$ is the global minimum, with magnitude 
\begin{equation}
    v^2 = \frac{4 \sqrt{3} M_0}{\Lambda_0 + \Lambda_1 + 3 \Lambda_2 -\sqrt{3}\Lambda_4}.
\end{equation}

The masses are given by the expressions
\begin{align}
    m_{h_i}^2 & = \frac{v^2}{24} \left( a + c + b + c - \sqrt{(a - c)^2 + (b - c)^2} \right) \notag  \qquad \textrm{(double degenerate)}\\
    m_{H_i}^2 & = \frac{v^2}{24} \left( a + c + b + c + \sqrt{(a - c)^2 + (b - c)^2} \right) \notag \qquad \textrm{(double degenerate)}\\
    m_{H_i^\pm} & = -\frac{1}{12} v^2 \left(3 \Lambda_1+3 \Lambda_2-2 \sqrt{3} \Lambda_4\right) \,, \qquad \frac{1}{12} v^2 \left(\sqrt{3} \Lambda_4 - 6 \Lambda _2\right)
\end{align}
where we defined
\begin{align}
    a = 3 (\Lambda_1 - \Lambda_2) & & b = 2(\Lambda_3 - \Lambda_1) & & c = \sqrt{3}\Lambda_4 \,.
\end{align}

\section{Softly-broken phenomenology \label{sec:softly}}

\subsection{Managing expectations}
The procedure for introducing symmetry breaking parameters in this way was described in \cite{deMedeirosVarzielas:2021zqs}. Here we summarize both it and its derivation, for convenience.

Given a potential $V_0$ with a quadratic term $V_2 = -m^2 \Phi_i^\dagger \Phi_i$ and an otherwise generic quartic part $V_4$, if one considers as independent variables $\Phi_i$ and $\Phi_i^\dagger$, the usual extremization conditions result in the set of equations

\begin{equation}
    \frac{\partial V_0}{\partial \Phi_i^*} = - m^2 \Phi_i + \frac{\partial V_4}{\partial \Phi_i^*} = 0 \,.
\end{equation}

\noindent implying that at any critical point we have the equality 

\begin{equation}
     \frac{\partial V_4}{\partial \Phi_i^*} = m^2 \Phi_i \,.
     \label{eq:equalsym}
\end{equation}

Now, we break the symmetry by considering the potential $V=V_0+V_{soft}$, with $V_{soft} = \Phi_i^\dagger M_{ij} \Phi_j$ and $M$ being an hermitian matrix.

The extremization conditions pick up extra terms, namely

\begin{equation}
    \frac{\partial V}{\partial \Phi_i^*} = M_{ij} \Phi_j - m^2 \Phi_i + \frac{\partial V_4}{\partial \Phi_i^*} = 0 \,.
    \label{eq:brokensym}
\end{equation}

\noindent The solutions to these equations are in the general case different from the fully symmetric model, and \eqref{eq:equalsym} ceases to be valid. However, if we now impose that a given vev alignment should remain a solution to these equations, up to a scaling factor, that is, $v|_{V} = \zeta \cdot v|_{V_0}$, we can nevertheless state that, at the scaled critical point,

\begin{equation}
    \frac{\partial V_4}{\partial \Phi_i^*} = \zeta^2 m^2 \Phi_i \,,
\end{equation}

\noindent due to the fact that $V_2$ and $V_4$ are homogeneous functions of degree 2 and 4, respectively. Introducing this in \eqref{eq:brokensym} we obtain that $M$ must satisfy

\begin{equation}
    M_{ij} \Phi_j = (1 - \zeta^2) m^2 \Phi_i \,.
\end{equation}

\noindent The upshot from this is that a given vev alignment is only preserved as a critical point if and only if it is an eigenvalue of the  soft-breaking parameter (SBP) matrix $M$. This gives us a simple way of constructing the most general matrix that preserves any given vev alignment:

\begin{equation}
\label{eq::matsbp}
    M = \mu_1 \overrightarrow{n_1} \overrightarrow{n_1}^\dagger + \mu_2 \overrightarrow{n_2} \overrightarrow{n_2}^\dagger + \mu_3 \overrightarrow{n_3} \overrightarrow{n_3}^\dagger \,.
\end{equation}

\noindent Within this formula, we take $\overrightarrow{n_1}$ as the desired vev alignment, and thus its eigenvalue is $\mu_1 = (1 - \zeta^2) m^2$. Regarding the remaining terms, we are free to choose any two eigenvalues and any two eigenvectors, as long as we guarantee that the set $\{\overrightarrow{n_1}, \overrightarrow{n_2}, \overrightarrow{n_3}\}$ is an orthonormal basis of $\mathbb{C}^3$. For that, one can start from two vectors $\{\overrightarrow{e_2}, \overrightarrow{e_3}\}$ that satisfy this condition and write the most general pair $\{\overrightarrow{n_2}, \overrightarrow{n_3}\}$ via a rotation:

\begin{equation}
    \begin{pmatrix}
      \overrightarrow{n_2}\\
      \overrightarrow{n_3}
    \end{pmatrix} = \begin{pmatrix}
      \cos \theta &e^{i \xi}\sin \theta \\
      -e^{-i \xi}\sin \theta & \cos \theta \\
    \end{pmatrix} \begin{pmatrix}
      \overrightarrow{e_2}\\
      \overrightarrow{e_3}
    \end{pmatrix}.
\end{equation}

\noindent This leaves us with a total of five degrees of freedom for parametrizing the SBPs, three eigenvalues and two angular variables. For the sake of simplicity, we now impose for the rest of this work $\mu_1=0$ and we are left with four parameters, which can assume any real value. By setting $\mu_1=0$ we are only imposing that the vev magnitude itself is also preserved (the alignment is preserved regardless of $\mu_1 \neq 0$).
However, it is possible to restrict $\theta$ to lie in the first quadrant without loss of generality, due to the fact that any $\{\overrightarrow{n_2}, \overrightarrow{n_3}\}$ that one obtains with $\theta \geq \pi/2$ can be brought to a configuration reachable by $0 \leq \theta < \pi/2$ by application of the appropriate combination of sign flips (under which $M$ is invariant) and $\overrightarrow{n_2} \leftrightarrow \overrightarrow{n_3}$, equivalent to $\mu_2 \leftrightarrow \mu_3$.
Thus, the minimal description of the SBPs is defined by the dimensionless parameters

\begin{align}
\label{eq::domain}
    \Sigma = - \frac{\sqrt{3}}{2} \frac{ \mu_2 + \mu_3 }{M_0} & & \Delta= - \frac{\sqrt{3}}{2} \frac{ \mu_2 - \mu_3 }{M_0} & & \xi & & 0 \leq \theta < \pi/2.
\end{align}

\noindent Here, $M_0$ is the coupling parameter of $V_2$. This recasting of parameters was done in order to simplify the results to be shown in the following sections.

Before proceeding, one important point regarding the ranges of admissible values for $\mu_2$ and $\mu_3$ needs to be addressed. From the form of \eqref{eq::matsbp} it is easily seen that if we choose $\mu_2$, $\mu_3>0$, then, for a given vev alignment, $M$ will not contribute to $V$ for this direction and give a positive contribution to any other direction, making our selected point deeper by comparison. Thus, if we start in a region of the $\Lambda$ parameter space that selects one of the four directions shown in \eqref{eq::symmvevs} and choose $\Sigma < 0$, $|\Delta| < \Sigma$ (equivalent to $\mu_2, \mu_3>0$) then we are guaranteed that the selected direction continues to be the global minimum, and as such these are sufficient conditions to ensure that the critical point is still the global minimum. Nevertheless, it is possible to find points outside of this region and/or that break the inequalities in the $\Lambda$ parameter space where our selected alignment continues to be the global minimum, if our deviations from them are not too large. Quantifying this magnitude analytically is not a straightforward task, however, and a numerical approach is better suited.
We emphasize then that $\mu_2, \mu_3>0$ are sufficient but not necessary conditions to have the selected direction remain a global minimum.

The upshot from this is that while it is always possible to choose a set of parameters that imposes a given direction from the fully-symmetric model as the global minimum, there is also some added freedom in the region of parameters, introduced by the SBPs, where one can explore new phenomenology.

\subsection{Softly-broken mass spectra}

Applying the procedure described above we studied the mass spectra predicted by the softly-broken $A_4$ potential (the spectra for the specific case of $S_4$ can be consulted in Appendix \ref{app:S4}).
Before proceeding to the full set of results, we provide an overview of some common properties displayed between the vev directions. Starting by the completely general properties, we have that:

\begin{itemize}
    \item Due to the way in which the symmetry was broken (and by setting $\mu_1 = 0$) the SM-like part of the spectrum behaves in exactly the same way as in the fully symmetric model, regarding both the expressions for $m_{H_{SM}}^2 = \frac{2}{\sqrt{3}} M_0$ and the scalar alignment, meaning that $H_{SM}$ is the only scalar that couples to the gauge bosons.
    \item All the mass degeneracies present in the parent model can be lifted by an appropriate choice of the soft breaking parameters.
\end{itemize}

Then, there are also properties shared among three out of the four alignments:

\begin{itemize}
    \item The mass splittings for the neutral bosons obey a specific pattern. Following the previously described nomenclature for the extended scalar sector, we have that
    \begin{equation}
    m_{H_2}^2 - m_{H_1}^2 = m_{h_2}^2 - m_{h_1}^2.
    \end{equation}
   \begin{center} \textit{Not observed for the vev alignment $(1, e^{i \alpha}, 0)$ in $A_4$ and $(1,i,0)$ in $S_4$.}
   \end{center}
    \item The mass expressions depend only on a specific combination of the angular parameters, effectively rendering them dependent on only three independent SBPs, meaning that one can travel along a certain direction in parameter space without affecting the values of the masses.
 \begin{center} \textit{Not observed for the vev alignment $(1, 0, 0)$ in $A_4$.} \end{center}
\end{itemize}

It is important to mention that all of these properties were also found in the $\Sigma(36)$ symmetric model \cite{deMedeirosVarzielas:2021zqs}. The fact that they do not hold for all the vev alignments in $A_4$/$S_4$ can be attributed to the fact that $\Sigma(36)$ is a more symmetric model than the ones we have at hand here.

Lastly, some remarks about the notation used are in order. The results are written in terms of a set of auxiliary parameters, $\Gamma_i$, which are alignment dependent linear combinations of the $\Lambda_i$, given below for each case. Through them, we always have that

\begin{equation}
    v^2 = \frac{2 \sqrt{3} M_0}{\Gamma_3},
    \label{eqn:generalvevparams}
\end{equation}

\noindent with $v^2 = v_1^2 + v_2^2 + v_3^2$.
We also make use of the following definitions, which hold for all alignments,

\begin{align}
\Gamma_{ij}^\pm & = \Gamma_i \pm \Gamma_j \\
x & = 2 \sin 2\theta \sin \xi
\label{eqn:generalparams}
\end{align}

\subsubsection*{$\mathbf{(1,0,0)}$:}
Here we chose

\begin{align}
      \overrightarrow{e_2} =
  \begin{pmatrix}
      0 \\
      1 \\
      0
    \end{pmatrix} & & \overrightarrow{e_3} =
  \begin{pmatrix}
      0 \\
      0 \\
      1
    \end{pmatrix} \,.
\end{align}

\noindent Then, with the alignment specific parameters

\begin{equation}
\begin{cases}
\Gamma_1 & = 3 \left( \Lambda_1 - \Lambda_3 \right) \\
\Gamma_2 & = 3 \left( \Lambda_2 - \Lambda_3 \right) \\
\Gamma_3 & = 2 \left( \Lambda_0 + \Lambda_3 \right) \\
\Gamma_4 & = 3 \Lambda_4 \,.
\end{cases} \,,
\end{equation}

\noindent the masses of the neutral bosons are given by

\begin{align}
m_{h_1}^2 & = \frac{v^2}{12} \left(\Gamma_{12}^+ - \Sigma \Gamma_3 - \sqrt{\left( \Gamma_{12}^- \right)^2 + \Delta^2 \Gamma_3^2 + \Gamma_4^2 + \left| \Delta \Gamma_3\right| s} \right) \notag \\
m_{h_2}^2 & = \frac{v^2}{12} \left(\Gamma_{12}^+ - \Sigma \Gamma_3 - \sqrt{\left( \Gamma_{12}^- \right)^2 + \Delta^2 \Gamma_3^2 + \Gamma_4^2 - \left| \Delta \Gamma_3\right| s} \right) \notag \\
m_{H_1}^2 & = \frac{v^2}{12} \left(\Gamma_{12}^+ - \Sigma \Gamma_3 + \sqrt{\left( \Gamma_{12}^- \right)^2 + \Delta^2 \Gamma_3^2 + \Gamma_4^2 - \left| \Delta \Gamma_3\right| s} \right) \notag \\
m_{H_2}^2 & = \frac{v^2}{12} \left(\Gamma_{12}^+ - \Sigma \Gamma_3 + \sqrt{\left( \Gamma_{12}^- \right)^2 + \Delta^2 \Gamma_3^2 + \Gamma_4^2 + \left| \Delta \Gamma_3\right| s} \right) \,,
\end{align}

\noindent with $s$ defined as

\begin{equation}
s = 2 \sqrt{\left( \Gamma_{12}^- \right)^2 (1 - s^2_ {\xi} s^2_{2\theta}) + \Gamma_4^2 (1 - c^2_{\xi} s^2_{ 2\theta}) - \Gamma_{12}^- \Gamma_4  s_{2\xi} s^2_{2\theta}} \,.
\end{equation}

\noindent Regarding the charged pairs, we have

\begin{align}
m_{H_1^\pm}^2 &= \frac{v^2}{6} \left(-3 \Lambda_3 - \frac{\Gamma_3}{2} (\Sigma + \Delta) \right) \implies \Delta m_{H_1^\pm}^2 = - \frac{\sqrt{3} M_0}{6} \mu_2 \notag \\
m_{H_2^\pm}^2 &= \frac{v^2}{6} \left(-3 \Lambda_3 - \frac{\Gamma_3}{2} (\Sigma - \Delta) \right) \implies \Delta m_{H_2^\pm}^2 = - \frac{\sqrt{3} M_0}{6} \mu_3 \,.
\label{eqn:charged100A4}
\end{align}

\noindent In this alignment, the charged splittings are simply linear with $\mu_2$ and $\mu_3$.

\subsubsection*{$\mathbf{(1,1,1)}$:}
We chose as orthogonal vectors

\begin{align}
      \overrightarrow{e_2} =
  \frac{1}{\sqrt{2}}\begin{pmatrix}
      1 \\
      -1 \\
      0
    \end{pmatrix} & & \overrightarrow{e_3} =
  \frac{1}{\sqrt{6}} \begin{pmatrix}
      1 \\
      1 \\
      -2
    \end{pmatrix} \,.
\end{align}
The alignment specific parameters are

\begin{equation}
\begin{cases}
    \Gamma_1 & = 2(\Lambda_3 - \Lambda_1) \\
    \Gamma_2 & = 3(\Lambda_2 - \Lambda_1) \\
    \Gamma_3 & = 2(\Lambda_0 + \Lambda_1) \\
    \Gamma_4 & = 2\sqrt{3}\Lambda_4
    \end{cases}
    \,.
\end{equation}

\noindent It is important to highlight the approximate invariance between this set of $\Gamma_i$ and the previous set under the interchange $\Lambda_1 \leftrightarrow \Lambda_3$. This approximate invariance was observed in the fully-symmetric model and can be seen as an inherited trait.

The masses for the neutral sector are
    \begin{align}
      m_{h_1}^2 & = \frac{v^2}{12} \left( \Gamma_{12}^+ - \Sigma \Gamma_3 - \sqrt{ \left( \Gamma_{12}^- \right)^2 + \Delta^2 \Gamma_3^2 + \Gamma_4^2 - \Delta \Gamma_3 \Gamma_4 x + \left| \Delta \Gamma_3 \Gamma_{12}^- \right| \sqrt{4 - x^2}} \right) \notag\\
      m_{h_2}^2 & = \frac{v^2}{12} \left( \Gamma_{12}^+ - \Sigma \Gamma_3 - \sqrt{ \left( \Gamma_{12}^- \right)^2 + \Delta^2 \Gamma_3^2 + \Gamma_4^2 - \Delta \Gamma_3 \Gamma_4 x - \left| \Delta \Gamma_3 \Gamma_{12}^- \right| \sqrt{4 - x^2}} \right) \notag \\
      m_{H_1}^2 & = \frac{v^2}{12} \left( \Gamma_{12}^+ - \Sigma \Gamma_3 + \sqrt{ \left( \Gamma_{12}^- \right)^2 + \Delta^2 \Gamma_3^2 + \Gamma_4^2 - \Delta \Gamma_3 \Gamma_4 x - \left| \Delta \Gamma_3 \Gamma_{12}^- \right| \sqrt{4 - x^2}} \right)  \notag\\
      m_{H_2}^2 & = \frac{v^2}{12} \left( \Gamma_{12}^+ - \Sigma \Gamma_3 + \sqrt{ \left( \Gamma_{12}^- \right)^2 + \Delta^2 \Gamma_3^2 + \Gamma_4^2 - \Delta \Gamma_3 \Gamma_4 x + \left| \Delta \Gamma_3 \Gamma_{12}^- \right| \sqrt{4 - x^2}} \right) \,,
    \end{align}

\noindent and for the charged scalars

    \begin{align}
      m_{H_1^\pm}^2 & = \frac{v^2}{12} \left( - 6 \Lambda_1 - \Sigma \Gamma_3 - \sqrt{\left( \frac{\Gamma_4}{2}\right)^2 + \Delta^2 \Gamma_3^2 - \Delta \frac{\Gamma_4}{2}\Gamma_3 x}\right) \notag \\
      m_{H_2^\pm}^2 & = \frac{v^2}{12} \left( - 6 \Lambda_1 - \Sigma \Gamma_3 + \sqrt{\left( \frac{\Gamma_4}{2}\right)^2 + \Delta^2 \Gamma_3^2 - \Delta \frac{\Gamma_4}{2}\Gamma_3 x}\right) \,.
    \end{align}

\subsubsection*{$\mathbf{(1, e^{i \alpha},0)}$:}

Our choice of basis is

\begin{align}
      \overrightarrow{e_2} =
  \frac{1}{\sqrt{2}}\begin{pmatrix}
      1 \\
      -e^{i\alpha} \\
      0
    \end{pmatrix} & & \overrightarrow{e_3} =
  \begin{pmatrix}
      0 \\
      0 \\
      1
    \end{pmatrix}.
\end{align}

In this case we were unable to obtain the analytical expressions for the masses of the neutral bosons, due to the added algebraic complexity introduced by the parameter dependent alignment. We turned to a numerical approach. We note however that the analytical results for the limiting case of $S_4$ were obtained and are discussed in Appendix \ref{app:S4}.

For the numeric analysis, we chose a benchmark set of parameters as

\begin{align}
\begin{cases}
M_0 = 10 \\
    \Sigma = -0.2 \\
    \Delta = 0.12 \\
\theta = \frac{\pi}{7} \\
\xi = \frac{\pi}{7}    
\end{cases} & & 
    \begin{cases}
    \Lambda_0 = 6 \\
    \Lambda_1 = 5 \\
\Lambda_2 = -4 \\
\Lambda_3 = 3 \\
\Lambda_4 = -0.3
    \end{cases}.
    \label{eq::params_num}
\end{align}

\noindent For the SBPs, \eqref{eq::params_num} should be interpreted as the values chosen when the relevant parameter is not the independent variable being analyzed, meaning that, for example, in Figure \ref{fig::delta}, $\theta = \xi = \pi/7$ and $\Delta$ varies in the range shown.

With this, the results are as follows:

\begin{figure}[H]
    \centering
    \resizebox{.6\linewidth}{!}{\input{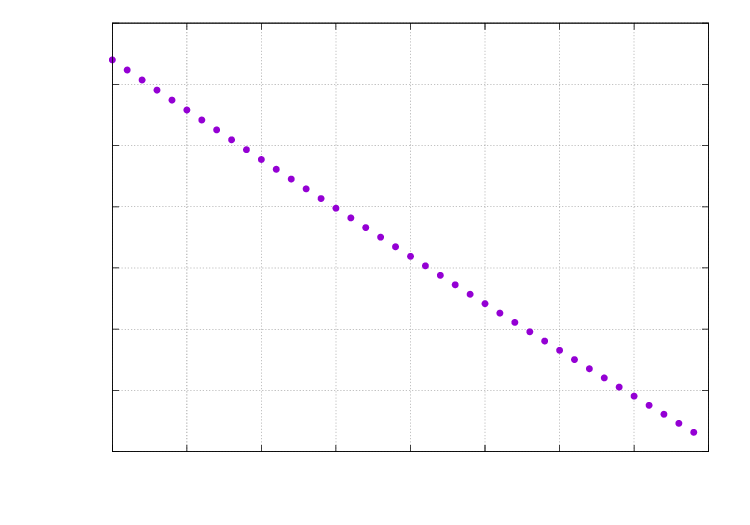}}
\caption{Numerical dependence of mass ratio with $\Delta$. \label{fig::delta}}
\end{figure}

\begin{figure}[H]
\begin{subfigure}[t]{0.48\columnwidth}
       \centering
    \resizebox{\columnwidth}{!}{\input{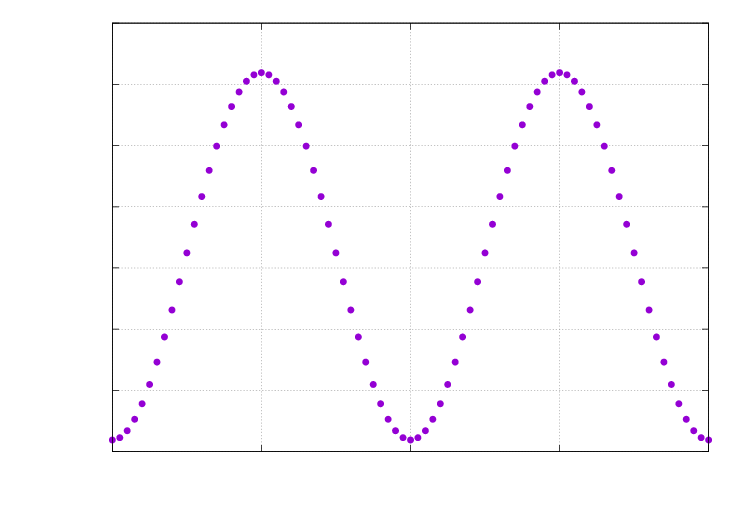}}
        \subcaption{}
        \end{subfigure}
        \hfill
        \begin{subfigure}[t]{0.48\columnwidth}
         \centering
    \resizebox{\columnwidth}{!}{\input{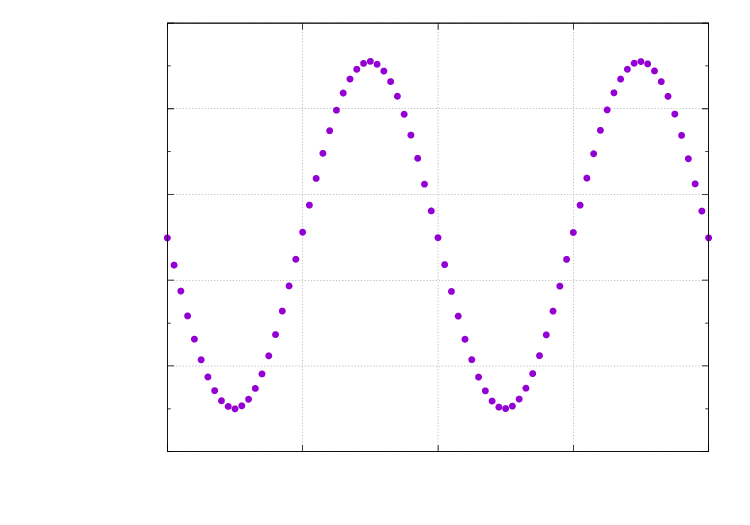}}
    \subcaption{}
        \end{subfigure}
     \caption{Numerical dependence of mass ratio with $\theta$ and $\xi$. \label{fig::a4masses}}
\end{figure}

Numerically, we were able to verify that (like the corresponding $S_4$ case) the dependence on $\Sigma$ is merely an overall shift of all the masses and therefore we omit its presentation here. Likewise, the dependence on $\Delta$ modulates the masses of the distinct eigenstates as a pre-factor as in $S_4$, see Figure \ref{fig::delta}. The dependence on $\theta$ and $\xi$ appears through $\cos 2\theta$ and $\sin 2\xi$ as seen in Figure \ref{fig::a4masses}. This last dependence is a new feature of $A_4$, and numerically we observe that the amplitude of the $\cos 2\theta$ dependence is considerably greater than that of the $\sin 2\xi$, as shown in the scales of the respective plots.

Regarding the charged scalars, defining the $\Gamma_i$ as

\begin{equation}
  \begin{cases}
  \Gamma_1 & = \frac{3}{4}\left(3(\Lambda_1 + \Lambda_2) + 2\Lambda_3 + 3(\Lambda_1 - \Lambda_2)\sec 2\alpha\right)\\
   \Gamma_2 & = \frac{1}{4}\left(3(\Lambda_1 + \Lambda_2) - 6\Lambda_3 + 3(\Lambda_1 - \Lambda_2)\sec 2\alpha\right)\\
 \Gamma_3 & = \frac{1}{4}\left(8\Lambda_0 + 2\Lambda_3 + 3(\Lambda_1 + \Lambda_2) + 3(\Lambda_1 - \Lambda_2)\sec 2\alpha\right)\\
  \end{cases} \,,
  \end{equation}

\noindent we have that the masses are given by

\begin{align}
    m_{H_1^\pm}^2 & = \frac{v^2}{12} \left( - \Gamma_{1} - \Sigma \Gamma_3 - \sqrt{\Gamma_{2}^2 + \Delta^2 \Gamma_3^2 + 2 \Delta \Gamma_{2}\Gamma_3 \cos 2\theta}\right) \notag \\
    m_{H_2^\pm}^2 & = \frac{v^2}{12} \left( - \Gamma_{1} - \Sigma \Gamma_3 + \sqrt{\Gamma_{2}^2 + \Delta^2 \Gamma_3^2 + 2 \Delta \Gamma_{2}\Gamma_3 \cos 2\theta}\right) \,.
  \end{align}

\subsubsection*{$\mathbf{(1, \omega,\omega^2)}$:}
For this alignment, we define $\omega = e^{2i\pi/3}$, and with this our basis vectors are

\begin{align}
      \overrightarrow{e_2} =
  \frac{1}{\sqrt{2}}\begin{pmatrix}
      1 \\
      -\omega \\
      0
    \end{pmatrix} & & \overrightarrow{e_3} =
  \frac{1}{\sqrt{6}} \begin{pmatrix}
      1 \\
      \omega \\
      -2\omega^2
    \end{pmatrix}.
\end{align}

The alignment specific parameters are

\begin{equation}
  \begin{cases}
  \Gamma_1 & = 3(\Lambda_1 - \Lambda_2)\\
  \Gamma_2 & = 2(\Lambda_3 - \Lambda_1)\\
  \Gamma_3 & = \frac{1}{2}\left(4 \Lambda_0 + \Lambda_1 + 3 \Lambda_2 - \sqrt{3} \Lambda_4\right) \\
  \Gamma_4 & = \sqrt{3}\Lambda_4
  \end{cases}
  \,,
  \end{equation}

\noindent through which one can write the masses of the neutral bosons as

  \begin{align}
    m_{h_1}^2 & = \frac{v^2}{12} \left( \Gamma_{14}^+ + \Gamma_{24}^+ - \Sigma \Gamma_3 - \sqrt{ \left( \Gamma_{14}^- \right)^2 + \left( \Gamma_{24}^- \right)^2 + \Delta^2 \Gamma_3^2 - \Delta \Gamma_3 \Gamma_{14}^- x + \left| \Delta \Gamma_3 \Gamma_{24}^- \right| \sqrt{4 - x^2}} \right) \notag \\
    m_{h_2}^2 & = \frac{v^2}{12} \left( \Gamma_{14}^+ + \Gamma_{24}^+ - \Sigma \Gamma_3 - \sqrt{ \left( \Gamma_{14}^- \right)^2 + \left( \Gamma_{24}^- \right)^2 + \Delta^2 \Gamma_3^2 - \Delta \Gamma_3 \Gamma_{14}^- x - \left| \Delta \Gamma_3 \Gamma_{24}^- \right| \sqrt{4 - x^2}} \right) \notag \\
    m_{H_1}^2 & = \frac{v^2}{12} \left( \Gamma_{14}^+ + \Gamma_{24}^+ - \Sigma \Gamma_3 + \sqrt{ \left( \Gamma_{14}^- \right)^2 + \left( \Gamma_{24}^- \right)^2 + \Delta^2 \Gamma_3^2 - \Delta \Gamma_3 \Gamma_{14}^- x - \left| \Delta \Gamma_3 \Gamma_{24}^- \right| \sqrt{4 - x^2}} \right) \notag \\
    m_{H_2}^2 & = \frac{v^2}{12} \left( \Gamma_{14}^+ + \Gamma_{24}^+ - \Sigma \Gamma_3 + \sqrt{ \left( \Gamma_{14}^- \right)^2 + \left( \Gamma_{24}^- \right)^2 + \Delta^2 \Gamma_3^2 - \Delta \Gamma_3 \Gamma_{14}^- x + \left| \Delta \Gamma_3 \Gamma_{24}^- \right| \sqrt{4 - x^2}} \right) \,,
  \end{align}

\noindent and the charged part of the spectrum as
  \begin{align}
    m_{H_1^\pm}^2 & = \frac{v^2}{12} \left( - 6 \Lambda_2 - \frac{\Gamma_{14}^-}{2} + \Gamma_4 - \Sigma \Gamma_3 - \sqrt{\left( \frac{\Gamma_{14}^-}{2}\right)^2 + \Delta^2 \Gamma_3^2 - \Delta \frac{\Gamma_{14}^-}{2}\Gamma_3 x}\right) \notag \\
    m_{H_2^\pm}^2 & = \frac{v^2}{12} \left( - 6 \Lambda_2 - \frac{\Gamma_{14}^-}{2} + \Gamma_4 - \Sigma \Gamma_3 + \sqrt{\left( \frac{\Gamma_{14}^-}{2}\right)^2 + \Delta^2 \Gamma_3^2 - \Delta \frac{\Gamma_{14}^-}{2}\Gamma_3 x}\right) \,.
  \end{align}

\subsection{Dark matter candidates}

In the $S_4$ and $A_4$ fully-symmetric models each vev alignment preserves a (different) subset of symmetries of the potential, leading to a scalar sector with stable neutral states and therefore to potential dark matter candidates stabilized by those residual symmetries. Now, in general, when one introduces $M$ into the potential some of the residual symmetries will be broken while others might still be conserved. If all the symmetries are broken then the previously stabilized fields can decay and the model ceases to have dark matter candidates. However, if one desires to retain this possibility then it is possible to look into the subset of symmetries preserved by the vev and see what conditions one must impose on $M$.

Thus, in this section, for each vev alignment we first determine if there are any symmetries left intact by the general form of \eqref{eq::matsbp}, and, if not, we determine the least restraining set of conditions on $M$ so that at least one symmetry is preserved. The analysis applies to $S_4$ and $A_4$, since the symmetry elements used are common to both. We verified all of our findings numerically.

\subsubsection*{$\mathbf{(1,0,0)}$:}

Here we focus on the symmetry element

\begin{equation}
\rho =
\begin{pmatrix}
1 &   &  \\
& -1 &  \\
&   & -1
\end{pmatrix},
\end{equation}

\noindent common to $A_4$ and $S_4$. This element generates a $Z_2$ subgroup, acting only on the second and third doublets by flipping their signs. It is straightforward to check that this symmetry is conserved by the vev direction $(1,0,0)$. Regarding the soft-breaking parameters, we have that $\rho$ acts on the contractions of the doublets as

\begin{align}
    \Phi_{1}^\dagger \Phi_2 \rightarrow - \Phi_{2}^\dagger \Phi_1 & & \Phi_{1}^\dagger \Phi_3 \rightarrow - \Phi_{1}^\dagger \Phi_3 \,,
\end{align}
\noindent and the remaining combinations are not shown since they map to themselves. These transformation properties force the elements of $M$ to satisfy

\begin{align}
M_{12} = - M_{21} \implies M_{12} = 0 & & M_{13} = - M_{31} \implies M_{13} = 0 \,, 
\end{align}

\noindent due to the hermiticity of $M$. Now, for every combination of the parameters $(\Sigma, \Delta, \theta, \xi)$ one always has that $M_{12}=M_{13}=0$, meaning that if one works in the region of parameters that selects this alignment as the global minimum all the SBP matrices generated by \eqref{eq::matsbp} are compatible with dark matter candidates, without adding further constraints. This feature contrasts with the findings in the $\Sigma(36)$ model \cite{deMedeirosVarzielas:2021zqs} where this vev direction led to a complete violation of the potential's symmetries (in $\Sigma(36)$ this $Z_2$ is not a symmetry of the potential).
It is interesting to consider how general is this feature, given this direction is a solution to potentials invariant under symmetries like $\Delta(54)$, where this same $Z_2$ given by $\rho$ is not a symmetry of the potential, but there are other $Z_2$ symmetries. Consider $\rho_{23}$:
\begin{equation}
\rho_{23} =
\begin{pmatrix}
1 &   &  \\
&  & 1 \\
&  1 & 
\end{pmatrix},
\end{equation}
which is a part of the $S_4$ and also the $\Delta(54)$ group in the chosen basis. This element is preserved by the vev. The potential with the respective SBPs does not remain invariant under $\rho_{23}$ (as $M_{22} \neq M_{33}$ in general).
The SBPs we are considering depend on the vev (and not on the original symmetry of the potential), so we conclude that potentials invariant under groups containing $\rho$ have dark matter candidates, whereas $\rho_{23}$ is not going to remain a residual symmetry due to the general SBPs.

\subsubsection*{$\mathbf{(1,e^{i\alpha},0)}$:}
In this case the fully-symmetric model only conserves two elements, which result from the composition of the $U(1)$ symmetry with a general $CP$ transformation and form two $Z_2$ subgroups:

\begin{equation}
\rho_{\pm} =
e^{i\alpha}\begin{pmatrix}
 & 1  &  \\
1 &  &  \\
&   & \pm 1
\end{pmatrix}_{CP} \,.
\end{equation}

\noindent Both of them are broken by the general form of $M$. To preserve either of these symmetries, and proceeding analogously to the previous case, we must have that 

\begin{align}
    M_{11}=M_{22} & & M_{23} = \pm M_{31} \,,
\end{align}

\noindent yielding the solutions

\begin{equation}
    \xi = \frac{\alpha - k\pi}{2}, \quad k \in \mathbb{Z} \,,
\end{equation}

\noindent which must be imposed as an additional constraint on the SBPs in order to have dark matter candidates.

\subsubsection*{$\mathbf{(1,1,1), (1,\omega,\omega^2)}$:}
Finally, we group these two directions due to their similar characteristics. In the fully-symmetric case both directions conserve multiple symmetries that are nevertheless completely broken by $M$.

In order to derive the least constraining relations we select the symmetry elements

\begin{align}
\rho_{13} =
\begin{pmatrix}
 &   & 1 \\
 & 1 &  \\
1 &  & 
\end{pmatrix}_{CP} & & 
\rho_{23} =
\begin{pmatrix}
1 &   &  \\
 &  & 1 \\
&  1 & 
\end{pmatrix}_{CP} \,,
\end{align}

\noindent which require that

\begin{align}
       \rho_{13}: \qquad M_{11} &= M_{33} & M_{12} &= M_{23} \\
       \rho_{23}: \qquad M_{22} &= M_{33} & M_{12} &= M_{31} \,,
\end{align}

\noindent whose solutions are

\begin{equation}
    \cos \xi = \pm \frac{\sqrt{3}}{\tan 2\theta} \,, \qquad \frac{\pi}{6} \leq \theta \leq \frac{\pi}{3} \,,
\end{equation}

\noindent which must be imposed as an additional constraint on the SBPs in order to have dark matter candidates.
Here, the positive solution corresponds to the condition obtained from $\rho_{23}$ and the negative solution from $\rho_{13}$. Regarding the bounds on $\theta$, they were obtained by combining the bounds on $\cos \xi$ with \eqref{eq::domain}.
\section{Conclusions \label{sec:con}}

We have studied softly-broken $A_4$ and $S_4$ 3 Higgs Doublet Models. We softly-break the symmetries in a controlled way such that the direction and magnitude of the symmetric vacuum expectations values are preserved. We present the mass spectra for each case after spontaneous symmetry breaking.

Having found cases with non-decaying mass eigenstates at leading order, we investigated in detail the residual symmetries left unbroken by the vacuum. In most cases these residual symmetries are broken by the soft-breaking terms and present the additional constraints required to preserve the residual symmetries. As an highlight, we have found one very appealing case where the most general vev-preserving soft-breaking terms preserve residual symmetries, without further constraints.

The residual symmetries, if left unbroken by both the vacuum and the soft-breaking parameters, lead to vanishing couplings between the different mass eigenstates and guarantee the stability of the lightest states charged under the symmetry. We conclude that in these cases there are stable dark matter candidates, which can occur in the softly-broken $A_4$ and $S_4$ 3 Higgs Doublet Models without additional symmetries being imposed.

\section{Acknowledgments}

We thank Igor Ivanov and Miguel Levy for helpful discussions.
IdMV acknowledges funding from Funda\c{c}\~{a}o para a Ci\^{e}ncia e a Tecnologia (FCT) through the contract IF/00816/2015 and was supported in part by FCT through projects CFTP-FCT Unit 777 (UID/FIS/00777/2019), PTDC/FIS-PAR/29436/2017, CERN/FIS-PAR/0004/2019 and CERN/FIS-PAR/0008/2019 which are partially funded through POCTI (FEDER), COMPETE, QREN and EU.
DI acknowledges funding from Funda\c{c}\~{a}o para a Ci\^{e}ncia e a Tecnologia (FCT) through project CERN/FIS-PAR/0008/2019.

\appendix

\section{Mass Spectra of $S_4$}
\label{app:S4}

\subsection{$(1,0,0), (1,1,1):$}
By setting $\Lambda_4 = 0$ in the expressions shown in the main text significant simplifications can be made to the results. In fact, for both of these directions the masses of the neutral bosons can be written through a set of universal formulas,

\begin{align}
	m_{h_1}^2 & =\frac{v^2}{12}\left(\Gamma_{12}^+ - \Sigma \Gamma_3 - \sqrt{(\Gamma_{12}^-)^2 + \Delta^2 \Gamma_3^2 + \left| \Delta \Gamma_{12}^- \Gamma_3 \right| \sqrt{4 - x^2}} \right) \notag \\
  m_{h_2}^2 & =\frac{v^2}{12}\left(\Gamma_{12}^+ - \Sigma \Gamma_3 - \sqrt{(\Gamma_{12}^-)^2 + \Delta^2 \Gamma_3^2 - \left| \Delta \Gamma_{12}^- \Gamma_3 \right| \sqrt{4 - x^2}} \right) \notag \\
  m_{H_1}^2 & =\frac{v^2}{12}\left(\Gamma_{12}^+ - \Sigma \Gamma_3 + \sqrt{(\Gamma_{12}^-)^2 + \Delta^2 \Gamma_3^2 - \left| \Delta \Gamma_{12}^- \Gamma_3 \right| \sqrt{4 - x^2}} \right) \notag \\
  m_{H_2}^2 & =\frac{v^2}{12}\left(\Gamma_{12}^+ - \Sigma \Gamma_3 + \sqrt{(\Gamma_{12}^-)^2 + \Delta^2 \Gamma_3^2 + \left| \Delta \Gamma_{12}^- \Gamma_3 \right| \sqrt{4 - x^2}} \right) \,,
	\label{eqn:masses1xx}
\end{align}
\noindent where the definitions of $x$ and $v^2$ are the same as in \eqref{eqn:generalvevparams} and \eqref{eqn:generalparams}.

Regarding the two pairs of charged Higgs, the difference to their masses in the unperturbed case is also universal between these two vev alignments,

\begin{align}
 	\Delta m_{H_1^\pm}^2 &= - \frac{\sqrt{3} M_0}{6} \left(\Sigma + \Delta \right)\notag \\
	\Delta m_{H_2^\pm}^2 &= - \frac{\sqrt{3} M_0}{6} \left(\Sigma - \Delta \right) \,.
\end{align}

\noindent While for $(1,0,0)$ this was already the case, it is a significant change for the case of $(1,1,1)$, where previously we had a different type of splitting.

The alignment specific parameters are now given by

\begin{align}
	(1,0,0): \begin{cases}
			\Gamma_1 = 3(\Lambda_1 - \Lambda_3) \\
			\Gamma_2 = 3(\Lambda_2 - \Lambda_3) \\
			\Gamma_3 = 2(\Lambda_0 + \Lambda_3)
		\end{cases} & \qquad \qquad \qquad (1,1,1): \begin{cases}
			\Gamma_1 = 2(\Lambda_3 - \Lambda_1) \\
			\Gamma_2 = 3(\Lambda_2 - \Lambda_1) \\
			\Gamma_3 = 2(\Lambda_0 + \Lambda_1)
		\end{cases} \,.
\label{eqn:gamma1xx}
\end{align}

Taken together, these results show that in $S_4$ these directions are extremely similar, with the only significant changes being the aforementioned interchange $\Lambda_1 \leftrightarrow \Lambda_3$ and the numerical factor in $\Gamma_1$.

\subsection{$(1,i,0)$:}
    In this case, the structure of the spectrum deviates from the previous two directions, and merits some discussion. 
    The expressions for the masses of the neutral bosons are
\begin{align}
  m_{h_1}^2 & = \frac{v^2}{12} \left( \Gamma_1 + \Gamma_2 - \Sigma \Gamma_3 - \sqrt{\left( \Gamma_1 - 3 \Gamma_2 \right)^2 + \Delta^2 \Gamma_3^2 + \Delta \left( \Gamma_1 - 3 \Gamma_2 \right) \Gamma_3 2 \cos 2\theta}\right) \notag\\
  m_{h_2}^2 & = \frac{v^2}{12} \left( 3\Gamma_1 - \Gamma_2 - \Sigma \Gamma_3 - \sqrt{\left( \Gamma_1 + \Gamma_2 \right)^2 + \Delta^2 \Gamma_3^2 - \Delta \left( \Gamma_1 + \Gamma_2 \right) \Gamma_3 2 \cos 2\theta}\right) \notag\\
  m_{H_1}^2 & = \frac{v^2}{12} \left( \Gamma_1 + \Gamma_2 - \Sigma \Gamma_3 + \sqrt{\left( \Gamma_1 - 3 \Gamma_2 \right)^2 + \Delta^2 \Gamma_3^2 + \Delta \left( \Gamma_1 - 3 \Gamma_2 \right) \Gamma_3 2 \cos 2\theta}\right) \notag\\
  m_{H_2}^2 & = \frac{v^2}{12} \left( 3\Gamma_1 - \Gamma_2 - \Sigma \Gamma_3 + \sqrt{\left( \Gamma_1 + \Gamma_2 \right)^2 + \Delta^2 \Gamma_3^2 - \Delta \left( \Gamma_1 + \Gamma_2 \right) \Gamma_3 2 \cos 2\theta}\right) \,,
  \label{eqn:masses1i0}
\end{align}

\noindent whereas the charged bosons have masses
\begin{align}
  m_{H_1^\pm}^2 & = \frac{v^2}{12} \left( - 6 \Lambda_2 - \Gamma_2 - \Sigma \Gamma_3 - \sqrt{\Gamma_2^2 + \Delta^2 \Gamma_3^2 - \Delta \Gamma_2 \Gamma_3 2 \cos 2 \theta}\right) \notag\\
  m_{H_2^\pm}^2 & = \frac{v^2}{12} \left( - 6 \Lambda_2 - \Gamma_2 - \Sigma \Gamma_3 + \sqrt{\Gamma_2^2 + \Delta^2 \Gamma_3^2 - \Delta \Gamma_2 \Gamma_3 2 \cos 2 \theta}\right) \,,
  \label{eqn:chargedmasses1i0}
\end{align}

\noindent with the following $\Gamma_i$
\begin{align}
\begin{cases}
  \Gamma_1 & = \frac{3}{2} \left( \Lambda_1 - \Lambda_2 \right) \\
  \Gamma_2 & = \frac{3}{2} \left( \Lambda_3 - \Lambda_2 \right) \\
  \Gamma_3 &= \frac{1}{2}\left(4 \Lambda_0 + \Lambda_3 + 3 \Lambda_2\right) \,.
\end{cases}
\end{align}

Focusing on the neutral bosons, one notices a different structure from all the remaining vev alignments, which leads to the curious situation when one sets $\sin2\theta=0$ there is always a degenerate pair, but the specific pair depends on the chosen parameters. More specifically, defining

\begin{align}
  a = \Gamma_1 + \Gamma_2 - \Delta \Gamma_3 & & b = \Gamma_1 - 3\Gamma_2 + \Delta \Gamma_3 \,,
\end{align}

\noindent if $(a>0, b>0)$ the degeneracy is between $(h_2, H_1)$, if $(a>0, b<0)$ the degeneracy is between $(h_1, h_2)$ and if $(a<0, b>0)$ the degeneracy is between $(H_1, H_2)$. The remaining case can not occur, since it would imply that $\Lambda_3 > \Lambda_1$, which cannot happen if we are to have $(1,i,0)$ as the global minimum (if we work in the region of parameter space allowed in the fully-symmetric model).

Finally, it is important to emphasize that this ``shifting'' degeneracy does not influence the ordering of the neutral bosonic masses if, again, one obeys the inequalities that select this vev as a global minimum in the fully-symmetric model.

\subsection{$(1,\omega,\omega^2)$:}
    This direction presents a spectrum more in line with the ones obtained from $(1,0,0)$ and $(1,1,1)$, but differing from them by having a more complicated combination of the $\Gamma_i$.
\begin{align}
    m_{h_1}^2 & =\frac{v^2}{12}\left(\Gamma_1 + \Gamma_2 - \Sigma \Gamma_3 - \sqrt{\Gamma_1^2 + \Gamma_2^2 + \Delta^2 \Gamma_3^2 - \Delta \Gamma_1 \Gamma_3 x + \left| \Delta \Gamma_2 \Gamma_3 \right| \sqrt{4 - x^2}} \right) \notag \\
  	m_{h_2}^2 & =\frac{v^2}{12}\left(\Gamma_1 + \Gamma_2 - \Sigma \Gamma_3 - \sqrt{\Gamma_1^2 + \Gamma_2^2 + \Delta^2 \Gamma_3^2 - \Delta \Gamma_1 \Gamma_3 x - \left| \Delta \Gamma_2 \Gamma_3 \right| \sqrt{4 - x^2}} \right) \notag \\
  	m_{H_1}^2 & =\frac{v^2}{12}\left(\Gamma_1 + \Gamma_2 - \Sigma \Gamma_3 + \sqrt{\Gamma_1^2 + \Gamma_2^2 + \Delta^2 \Gamma_3^2 - \Delta \Gamma_1 \Gamma_3 x - \left| \Delta \Gamma_2 \Gamma_3 \right| \sqrt{4 - x^2}} \right) \notag \\
  	m_{H_2}^2 & =\frac{v^2}{12}\left(\Gamma_1 + \Gamma_2 - \Sigma \Gamma_3 + \sqrt{\Gamma_1^2 + \Gamma_2^2 + \Delta^2 \Gamma_3^2 - \Delta \Gamma_1 \Gamma_3 x + \left| \Delta \Gamma_2 \Gamma_3 \right| \sqrt{4 - x^2}} \right) \,, 
\end{align}

\begin{align}
  m_{H_1^\pm}^2 &= \frac{v^2}{12} \left( -6 \Lambda_2 - \frac{\Gamma_1}{2} - \Sigma \Gamma_3 - \sqrt{\left(\frac{\Gamma_1}{2}\right)^2 + \Delta^2 \Gamma_3^2 - \Delta \frac{\Gamma_1}{2} \Gamma_3 x}\right) \notag \\
m_{H_2^\pm}^2 &= \frac{v^2}{12} \left( -6 \Lambda_2 - \frac{\Gamma_1}{2} - \Sigma \Gamma_3 + \sqrt{\left(\frac{\Gamma_1}{2}\right)^2 + \Delta^2 \Gamma_3^2 - \Delta \frac{\Gamma_1}{2} \Gamma_3 x}\right) \,,
\end{align}

\noindent where the $\Gamma_i$ are now defined as

\begin{align}
\begin{cases}
    \Gamma_1 &= 3\left(\Lambda_1 - \Lambda_2\right) \\
  \Gamma_2 &= 2\left(\Lambda_3 - \Lambda_1\right) \\
  \Gamma_3 &= \frac{1}{2}\left(4 \Lambda_0 + \Lambda_1 + 3 \Lambda_2\right) \,.
  \end{cases}
  \label{eqn:gamma1ww2}
\end{align}

\section{SBP Matrices}

In this Appendix we explicitly show the expressions for matrix $M$ described in \eqref{eq::matsbp} for the 4 vev alignments discussed in the main text for $A_4$, for the sake of completeness. Once again, the results for $S_4$ are obtained by setting $\Lambda_4=0$, which in the context of what follows means that $\alpha = \pi/2$. All expressions assume that $\mu_1=0$.

\subsection{$(1,0,0)$:}

For this alignment, with the choice 

\begin{align}
      \overrightarrow{e_2} =
  \begin{pmatrix}
      0 \\
      1 \\
      0
    \end{pmatrix} & & \overrightarrow{e_3} =
  \begin{pmatrix}
      0 \\
      0 \\
      1
    \end{pmatrix} \,,
\end{align}

\noindent we obtain that the entries of $M$ are as follows

\begin{align}
 M_{11} & = M_{21} = M_{31} = 0 \notag \\ 
 M_{22} & =\frac{1}{2} \left( \Sigma + \Delta  \cos 2 \theta\right) \notag \\
 M_{32} & = \frac{1}{2} e^{i \xi } \Delta \sin 2\theta \notag \\
 M_{33} & = \frac{1}{2} \left(\Sigma - \Delta  \cos 2 \theta \right).
\end{align}

\noindent The omitted terms are obtained from the hermiticity of $M$.

\subsection{$(1,1,1)$:}
With the choice of basis

\begin{align}
      \overrightarrow{e_2} =
  \frac{1}{\sqrt{2}}\begin{pmatrix}
      1 \\
      -1 \\
      0
    \end{pmatrix} & & \overrightarrow{e_3} =
  \frac{1}{\sqrt{6}} \begin{pmatrix}
      1 \\
      1 \\
      -2
    \end{pmatrix} \,,
\end{align}

\noindent one obtains that

\begin{align}
\label{eq::sbp111}
 M_{11} & = \frac{1}{6} \left(2 \Sigma +  \Delta (\cos 2 \theta + \sqrt{3} \sin 2 \theta \cos \xi) \right) \notag \\
 M_{21} & = \frac{1}{6} \left(\Delta (i \sqrt{3}  \sin 2 \theta  \sin \xi - 2 \cos 2 \theta) - \Sigma \right) \notag \\
 M_{22} & = \frac{1}{6} \left(2 \Sigma +  \Delta (\cos 2 \theta - \sqrt{3} \sin 2 \theta \cos \xi) \right) \notag \\
 M_{31} & =\frac{1}{6} \left( \Delta (\cos 2 \theta - \sqrt{3} e^{i \xi } \sin 2 \theta) - \Sigma \right) \notag \\
 M_{32} & =\frac{1}{6} \left( \Delta (\cos 2 \theta + \sqrt{3} e^{i \xi } \sin 2 \theta) - \Sigma \right) \notag \\
 M_{33} & =\frac{1}{3} (\Sigma -\Delta  \cos (2 \theta )).
\end{align}

\subsection{$(1,\omega,\omega^2)$:}

For this alignment, with the choice

\begin{align}
      \overrightarrow{e_2} =
  \frac{1}{\sqrt{2}}\begin{pmatrix}
      1 \\
      -\omega \\
      0
    \end{pmatrix} & & \overrightarrow{e_3} =
  \frac{1}{\sqrt{6}} \begin{pmatrix}
      1 \\
      \omega \\
      -2\omega^2
    \end{pmatrix},
\end{align}

\noindent one obtains the elements of $M$ by multiplying the expressions in \eqref{eq::sbp111} by the appropriate powers of $\omega$. Explicitly,

\begin{align}
    M_{ii} & = M_{ii}^{(1,1,1)} \qquad (i=1,2,3) \qquad (\textrm{no sum}) \notag \\
    M_{21} & = \omega M_{21}^{(1,1,1)} \notag \\
    M_{31} & = \omega^2 M_{31}^{(1,1,1)} \notag \\
    M_{32} & = \omega M_{32}^{(1,1,1)},
\end{align}

\noindent where the superscript $(1,1,1)$ indicates the corresponding matrix element for that alignment.

\subsection{$(1,e^{i \alpha},0)$:}

With the choice of basis of

\begin{align}
      \overrightarrow{e_2} =
  \frac{1}{\sqrt{2}}\begin{pmatrix}
      1 \\
      -e^{i\alpha} \\
      0
    \end{pmatrix} & & \overrightarrow{e_3} =
  \begin{pmatrix}
      0 \\
      0 \\
      1
    \end{pmatrix},
\end{align}

\noindent the matrix element are given by 

\begin{align}
 M_{11} & = M_{22} =  \frac{1}{4} (\Sigma + \Delta  \cos 2 \theta) \notag \\
 M_{21} & = -\frac{1}{4} e^{i \alpha} (\Sigma + \Delta  \cos 2 \theta ) \notag \\
 M_{31} & = -\frac{\Delta  e^{i \xi } \sin 2\theta }{2 \sqrt{2}} \notag \\ 
 M_{32} & = \frac{\Delta e^{i (\xi - \alpha)} \sin 2 \theta}{2\sqrt{2}} \notag \\
 M_{33} & = \frac{1}{2} (\Sigma - \Delta  \cos 2 \theta).
\end{align}

\end{document}